\documentclass[twocolumn,showpacs,preprintnumbers,amsmath,amssymb]{revtex4}

\def\be{\begin{equation}}
\def\ee{\end{equation}}

\def\be{\begin{equation}}
\def\ee{\end{equation}}

\def\fr{\frac}

\def\be{\begin{equation}}
\def\ee{\end{equation}}

\def\fr{\frac}

\def\De{\Delta}

\usepackage[dvipdfmx]{graphicx}
\usepackage{comment}
\usepackage{epsf}
\usepackage{amsmath}
\usepackage{graphics}
\usepackage{amsfonts}
\usepackage{amssymb}
\usepackage{latexsym}
\usepackage{color}
\input{colordvi.tex}
\usepackage{feynmp}

\begin{document}
\preprint{RESCEU-7/14}

\title{Reheating the Universe Once More: \\
The Dissipation of Acoustic Waves as a Novel Probe of Primordial Inhomogeneities\\
on Even Smaller Scales}


\author{Tomohiro Nakama$^{1,2}$}
\author{Teruaki Suyama$^{2}$}
\author{Jun'ichi Yokoyama$^{2,3}$}
\affiliation{
$^1$
  Department of Physics, Graduate School of Science,\\ The University of Tokyo, Tokyo 113-0033, Japan
}
\affiliation{
$^2$
  Research Center for the Early Universe (RESCEU), Graduate School
  of Science,\\ The University of Tokyo, Tokyo 113-0033, Japan
}
\affiliation{
$^3$
  Kavli Institute for the Physics and Mathematics 
of the Universe (Kavli IPMU), WPI, TODIAS,\\
The University of Tokyo, Kashiwa, Chiba 277-8568, Japan
}

\date{\today}

\begin{abstract}
We provide a simple but robust bound on the primordial curvature perturbation 
in the range $10^4 {\rm Mpc}^{-1} < k < 10^5 {\rm Mpc}^{-1}$, which has
 not been constrained so far unlike low wavenumber modes.
Perturbations on these scales dissipate the energy of their acoustic oscillations
by the Silk damping after primordial nucleosynthesis but before the redshift 
$z \sim 2 \times 10^6$ and 
reheat the photon bath without invoking CMB distortions.
This {\it acoustic reheating} results in the decrease of the 
baryon-photon ratio.
By combining independent measurements probing the nucleosynthesis era 
and around the recombination epoch, 
we find an upper bound on the amplitude of the curvature 
perturbation over the above wavenumber range as ${\cal P}_\zeta < 0.06$.
Implications for super massive black holes are also discussed.
\end{abstract}


\maketitle

\textit{Introduction.}
Primordial inhomogeneities have been intensively investigated by cosmic microwave background (CMB) \cite{Hinshaw:2012aka,Ade:2013uln} or 
large scale structures of the Universe. 
However, the perturbation scales relevant to these probes are limited to 
$\cal{O}(\mathrm{Mpc})$ to $\cal{O}(\mathrm{Gpc})$ and information of fluctuations on smaller scales 
is relatively scarce. 
On the other hand, 
some models of the early universe predict enhancement of the power spectrum of fluctuations on 
small scales 
\cite{PhysRevD.54.6040,
PhysRevD.42.3329,PhysRevD.50.7173,yokoyama-1997-673,
PhysRevD.58.083510,Jun'ichi1998133,kawasaki-1999-59,
PTPS.136.338,1475-7516-2008-06-024,PhysRevD.59.103505,
PhysRevD.63.123503,
PhysRevD.64.021301,Kawasaki:2007zz,Kawaguchi:2007fz},
 so investigating small scale perturbations is important. 
Given this situation, 
several methods to probe small scale fluctuations have been studied such as 
primordial black holes (PBHs) \cite{Bugaev:2008gw,Josan:2009qn}, 
ultracompact minihalos \cite{Bringmann:2011ut,Josan:2010vn,Li:2012qha,Scott:2012kx,Yang:2013dsa}, and
CMB spectral distortions 
\cite{Barrow2,Chluba:2011hw,Chluba:2012we,Chluba:2012gq,Dent:2012ne,Khatri:2012tw,Sunyaev:2013aoa,Khatri:2013dha,Chluba:2013dna,Chluba:2013pya}. 

In this \textit{Letter}, we propose a novel method to probe perturbations on 
smaller scales than those probed by CMB spectral distortions ($10^4\mathrm{Mpc}^{-1}<k$). 
This method is based on a phenomenon we call ``acoustic reheating". 
During the radiation-dominated era, perturbations are damped after the horizon crossing (diffusion damping or Silk damping \cite{Silk:1967kq,HSato}), 
injecting energy into the background universe. 
Before the $\mu$-era, or the epoch when energy release leads to $\mu$-distortions, 
any energy injection only causes increase in the average photon temperature, without 
invoking any spectral distortions \cite{Sunyaev:1970er,Burigana,Hu:1992dc}. 
If this energy injection takes place after the 
Big Bang Nucleosynthesis (BBN), it increases the number density 
of photons $n_{\gamma}$, without changing the number density of baryons $n_\mathrm{b}$, 
and so decreases the baryon-photon ratio $\eta\equiv n_{\mathrm{b}}/n_{\gamma}$. 
Since the value of $\eta$ is independently inferred by BBN \cite{Steigman:2014pfa} and CMB observation \cite{Ade:2013zuv}, 
we can put constraints on 
the amount of energy injection \cite{Simha:2008zj}, or primordial perturbation amplitude (see also \cite{Barrow1}).

\textit{Calculation of energy injection.}
The basic equations can be found in \cite{Chluba:2012we} (hereafter CEB). 
The total energy release due to the damping of acoustic waves from the redshift $z_2$ to $z_1(<z_2)$
is given by 
\be
\fr{\De \rho_{\gamma}}{\rho_{\gamma}}
=\int_{z_1}^{z_2}\fr{1}{a^4\rho_{\gamma}}\fr{d(a^4Q_{\mathrm{ac}})}{dz}dz,
\ee
with
\be
\fr{1}{a^4\rho_{\gamma}}\fr{d(a^4Q_{\mathrm{ac}})}{dz}
\sim 9.4a\int\fr{kdk}{k_{\mathrm{D}}^2}{\cal P}_{\zeta}(k)
2\sin^2(kr_{\mathrm{s}})e^{-2k^2/k_{\mathrm{D}}^2}, \label{dQ}
\ee
where $r_{\mathrm{s}}\sim 2.7\times 10^5(1+z)^{-1}$Mpc is the sound horizon and 
$k_{\mathrm{D}}\sim 4.0\times 10^{-6}(1+z)^{3/2}\mathrm{Mpc}^{-1}$ is the damping scale 
determined by the diffusion of photons. 

The largest contributions to the energy release at a redshift 
$z$ come from the modes around $k\sim k_{\mathrm{D}}(z)$ 
and so we can safely approximate $\sin^2(kr_{\mathrm{s}})\sim 1/2$, 
since $k_{\mathrm{D}}(z)r_{\mathrm{s}}(z)\sim (1+z)^{1/2}\gg 1$. 
Let us consider a top-hat power spectrum: 
${\cal P}_{\zeta}(k)=A_{\zeta}(k_1<k<k_2), 0(\mathrm{otherwise})$, 
noting that the effects of acoustic reheating are most significant when the width of the 
enhanced part of the power spectrum is widest. 
We set $k_1=10^4\mathrm{Mpc}^{-1}$, since 
the power spectrum is severely constrained for 
$k_1<10^4\mathrm{Mpc}^{-1}$ by $\mu$-distortion \cite{Chluba:2012we}. 
On the other hand, the modes $10^5\mathrm{Mpc}^{-1}<k$ dissipate before the neutrino decoupling 
due to the neutrino diffusion.
The comoving wave number for the neutrino diffusion becomes $k=10^5\mathrm{Mpc}^{-1}$ at
the time of neutrino decoupling, which is close to the horizon scale at that time \cite{Jedamzik:1996wp}. 
Since what can be probed by the consistency between BBN and CMB 
is only energy injection \textit{after} BBN, taking place shortly after the neutrino decoupling, 
modes shorter than $k=10^5~\mathrm{Mpc}^{-1}$ cannot be constrained and so 
we set $k_2=10^5\mathrm{Mpc}^{-1}$. 
Correspondingly, we choose $z_1=2\times 10^6$, the onset of $\mu$-era, 
and $z_2=8.5\times 10^6$, around when the mode $k=10^5~\mathrm{Mpc}^{-1}$ dissipates, assuming it 
dissipates due to the diffusion of photons. 

For the top-hat power spectrum ranging from $k_1$ to $k_2$, 
the energy release given by Eq.~(\ref{dQ}) becomes 
\begin{eqnarray}
\fr{1}{a^4\rho_{\gamma}}\fr{d(a^4Q_{\mathrm{ac}})}{dz}
\sim 
\fr{2.4A_{\zeta}}{1+z}
\left[
\exp\left\{-2\left(\fr{1+z_{*,1}}{1+z}\right)^3\right\}
\right.
\nonumber\\
\left.
-\exp\left\{-2\left(\fr{1+z_{*,2}}{1+z}\right)^3\right\}
\right],\label{Qac}
\end{eqnarray}
where $k_*\equiv 4\times 10^{-6}\mathrm{Mpc}^{-1}$ and
\be
\quad  z_{*,i}\equiv \left(\fr{k_i}{k_*}\right)^{2/3}(i=1,2)\label{kszs}
\ee
is the redshift when the mode $k_i$ dissipates. 
The total energy release then becomes 
\begin{eqnarray}
\fr{\De \rho_{\gamma}}{\rho_{\gamma}}
\sim 
0.8A_{\zeta}
\left[
\mathrm{Ei}
\left(
-2\left(
\fr{1+z_{*,1}}{1+z}
\right)^3
\right)\right.\nonumber\\
-\left.\mathrm{Ei}\left(-2\left(\fr{1+z_{*,2}}{1+z}\right)^3\right)
\right]^{z_1}_{z_2}\sim 2.3A_{\zeta},\label{deltarho}
\end{eqnarray}
where Ei denotes an exponential integral. 

\textit{Constraints on $A_{\zeta}$ obtained by the baryon-photon ratio.}
The baryon-photon ratio $\eta$ has been determined independently by BBN and CMB, so 
the damping should not increase the number of photons too much (or equivalently should not decrease $\eta$ too much) 
after BBN, 
from which constraints on $A_\zeta$ can be obtained. 
To be consistent with observation, we require (noting $n_\gamma \propto T^3, \rho_\gamma\propto T^4$)
\be
\fr{\eta_{\mathrm{CMB}}}{\eta_{\mathrm{BBN}}}
=\left(1-\fr{3}{4}\fr{\Delta \rho_{\gamma}}{\rho_{\gamma}}\right)
>\fr{\eta_{\mathrm{CMB,obs}}}{\eta_{\mathrm{BBN,obs}}},
\ee
where $\eta_{\mathrm{BBN}}$ and $\eta_{\mathrm{CMB}}$ are the baryon-photon ratios at the time of BBN and 
after the onset of the $\mu$-era ($\eta$ becomes constant after this moment since we only consider energy injection 
before the $\mu$-era); the subscript ``obs" implies a value determined by observation. 
Using (\ref{deltarho}), this inequality is rewritten as a constraint on $A_{\zeta}$:
\be
A_\zeta\lesssim 0.6\left(1-\fr{\eta_{\mathrm{CMB,obs}}}{\eta_{\mathrm{BBN,obs}}}\right).
\ee
As the observed values, 
we follow \cite{Steigman:2014pfa}, in which 
$\eta=(6.11\pm 0.08)\times 10^{-10}$ for CMB and $\eta=(6.19\pm 0.21)\times 10^{-10}$ for BBN were adopted. 
For 1$\sigma$ constraint, we conservatively set $\eta_{\mathrm{CMB,obs}}=(6.11-0.08)\times 10^{-10}$ and 
$\eta_{\mathrm{BBN,obs}}=(6.19+0.21)\times 10^{-10}$ (and 2$\sigma$ constraints are considered similarly). 
Then, the constraint on $A_{\zeta}$ is 
\be
A_\zeta\lesssim 0.03(1\sigma), \quad 0.06(2\sigma).
\ee
\textit{Discussion.}
The constraints on the amplitude of the curvature perturbation have also been 
obtained to avoid overproduction of PBHs to be consistent with 
observations \cite{Saito:2008jc,Saito:2009jt,Carr:2009jm}. 
If we follow \cite{Josan:2009qn} (see also \cite{Bugaev:2008gw}), 
considering the disruption of wide binaries, 
which is relevant to the scales accessible by acoustic reheating, 
we  obtain a constraint by PBHs as 
$A_{\zeta}\lesssim 0.05$.

Although the order of magnitude of these constraints is the same, we
may not compare the two directly for several reasons.
First a constraint imposed by PBH refers to the average of peaked
curvature fluctuations over one e-fold of wavenumber, and it is
obtained under the assumption that there is one-to-one correspondence
between the mass of PBHs and the comoving scale of perturbation.  But this
is not true since the critical phenomenon
\cite{PhysRevD.59.124013,Musco:2008hv} has been observed, which results
in formation of a number of PBHs with their mass much smaller than the
comoving horizon mass.  Furthermore, since PBHs are created at
high-$\sigma$ peaks, possible non-Gaussian distribution may change their
abundance in a model dependent manner
\cite{PhysRevD.54.6040,PhysRevD.50.7173,1475-7516-2008-06-024}.  
In particular, when non-Gaussianity is extremely large, 
it can change constraints to avoid 
overproduction of PBHs \cite{Byrnes:2012yx,Young:2013oia}. 
(Note that $f_{NL}^{local}={\cal O}(1)$ corresponds to a case with
extremely large non-Gaussianity for PBH formation whose relevant
amplitude of fluctuation is ${\cal O}(0.1)$ and the ratio of   linear-to- 
second-order term is as large as 0.1 for  $f_{NL}^{local}={\cal O}(1)$.)

On the other hand, acoustic reheating considered here 
is insensitive to the assumption of Gaussianity (as is also pointed out in CEB) 
and is relatively easy to quantify precisely as well as relate to observations. 
Furthermore, what is interesting about constraints on the amplitude of primordial fluctuations 
obtained by acoustic reheating is that 
they can improve almost in proportion to potential future decrease in 
error bars associated with the determination of $\eta$. 

Though our constraints apply only in a relatively narrow range $10^4\mathrm{Mpc}^{-1}<k<10^5\mathrm{Mpc}^{-1}$, 
this technique will have profound implications. 
For example, if the constraints from acoustic reheating become tighter in future, 
PBHs in the corresponding comoving horizon mass range $10^3M_\odot<M<10^5M_\odot$ will be severely constrained 
(note that PBHs bigger than $10^5M_\odot$ are severely constrained by $\mu$-distortion \cite{KNS}). 
This mass range is particularly interesting in the context of scenarios of 
PBHs as the seeds of super massive black holes.

\textit{Acknowledgements.}
This work was partially supported by JSPS Grant-in-Aid for Scientific Research
23340058 (J.Y.), Grant-in-Aid for Scientific Research on Innovative Areas No.
24103006 (J.Y.)  and No. 25103505 (T.S.), 
and Grant-in-Aid for JSPS Fellow No. 25.8199 (T.N.). 

\textit{Note added.}
During the final stages of preparing the manuscript we became
aware of work by Jeong, Pradler, Chluba and Kamionkowski, who also observed
the importance of reheating from second-order perturbations
focusing, however, on different aspects of BBN constraints and obtaining 
different results \cite{Jeong:2014gna}. 
We are grateful to the authors of \cite{Jeong:2014gna} for useful
communications in revising our manuscript.

\bibliography{ref.bib}

\begin{thebibliography}{52}
\expandafter\ifx\csname natexlab\endcsname\relax\def\natexlab#1{#1}\fi
\expandafter\ifx\csname bibnamefont\endcsname\relax
  \def\bibnamefont#1{#1}\fi
\expandafter\ifx\csname bibfnamefont\endcsname\relax
  \def\bibfnamefont#1{#1}\fi
\expandafter\ifx\csname citenamefont\endcsname\relax
  \def\citenamefont#1{#1}\fi
\expandafter\ifx\csname url\endcsname\relax
  \def\url#1{\texttt{#1}}\fi
\expandafter\ifx\csname urlprefix\endcsname\relax\def\urlprefix{URL }\fi
\providecommand{\bibinfo}[2]{#2}
\providecommand{\eprint}[2][]{\url{#2}}

\bibitem[{\citenamefont{Hinshaw et~al.}(2012)}]{Hinshaw:2012aka}
\bibinfo{author}{\bibfnamefont{G.}~\bibnamefont{Hinshaw}} \bibnamefont{et~al.}
  (\bibinfo{collaboration}{WMAP Collaboration}) (\bibinfo{year}{2012}),
  \eprint{1212.5226}.

\bibitem[{\citenamefont{Ade et~al.}(2013{\natexlab{a}})}]{Ade:2013uln}
\bibinfo{author}{\bibfnamefont{P.}~\bibnamefont{Ade}} \bibnamefont{et~al.}
  (\bibinfo{collaboration}{Planck Collaboration})
  (\bibinfo{year}{2013}{\natexlab{a}}), \eprint{1303.5082}.

\bibitem[{\citenamefont{Garcia-Bellido
  et~al.}(1996)\citenamefont{Garcia-Bellido, Linde, and
  Wands}}]{PhysRevD.54.6040}
\bibinfo{author}{\bibfnamefont{J.}~\bibnamefont{Garcia-Bellido}},
  \bibinfo{author}{\bibfnamefont{A.}~\bibnamefont{Linde}}, \bibnamefont{and}
  \bibinfo{author}{\bibfnamefont{D.}~\bibnamefont{Wands}},
  \bibinfo{journal}{Phys. Rev. D} \textbf{\bibinfo{volume}{54}},
  \bibinfo{pages}{6040} (\bibinfo{year}{1996}).

\bibitem[{\citenamefont{Hodges and Blumenthal}(1990)}]{PhysRevD.42.3329}
\bibinfo{author}{\bibfnamefont{H.~M.} \bibnamefont{Hodges}} \bibnamefont{and}
  \bibinfo{author}{\bibfnamefont{G.~R.} \bibnamefont{Blumenthal}},
  \bibinfo{journal}{Phys. Rev. D} \textbf{\bibinfo{volume}{42}},
  \bibinfo{pages}{3329} (\bibinfo{year}{1990}).

\bibitem[{\citenamefont{Ivanov et~al.}(1994)\citenamefont{Ivanov, Naselsky, and
  Novikov}}]{PhysRevD.50.7173}
\bibinfo{author}{\bibfnamefont{P.}~\bibnamefont{Ivanov}},
  \bibinfo{author}{\bibfnamefont{P.}~\bibnamefont{Naselsky}}, \bibnamefont{and}
  \bibinfo{author}{\bibfnamefont{I.}~\bibnamefont{Novikov}},
  \bibinfo{journal}{Phys. Rev. D} \textbf{\bibinfo{volume}{50}},
  \bibinfo{pages}{7173} (\bibinfo{year}{1994}).

\bibitem[{\citenamefont{Yokoyama}(1997)}]{yokoyama-1997-673}
\bibinfo{author}{\bibfnamefont{J.}~\bibnamefont{Yokoyama}},
  \bibinfo{journal}{Astron.\ Astrophys.} \textbf{\bibinfo{volume}{318:673}}
  (\bibinfo{year}{1997}).

\bibitem[{\citenamefont{Yokoyama}(1998{\natexlab{a}})}]{PhysRevD.58.083510}
\bibinfo{author}{\bibfnamefont{J.}~\bibnamefont{Yokoyama}},
  \bibinfo{journal}{Phys.\ Rev.\ D} \textbf{\bibinfo{volume}{58}},
  \bibinfo{pages}{083510} (\bibinfo{year}{1998}{\natexlab{a}}).

\bibitem[{\citenamefont{Yokoyama}(1998{\natexlab{b}})}]{Jun'ichi1998133}
\bibinfo{author}{\bibfnamefont{J.}~\bibnamefont{Yokoyama}},
  \bibinfo{journal}{Phys. Rep.} \textbf{\bibinfo{volume}{307}},
  \bibinfo{pages}{133 } (\bibinfo{year}{1998}{\natexlab{b}}).

\bibitem[{\citenamefont{Kawasaki and Yanagida}(1999)}]{kawasaki-1999-59}
\bibinfo{author}{\bibfnamefont{M.}~\bibnamefont{Kawasaki}} \bibnamefont{and}
  \bibinfo{author}{\bibfnamefont{T.}~\bibnamefont{Yanagida}},
  \bibinfo{journal}{Phys.\ Rev.\ D} \textbf{\bibinfo{volume}{59}},
  \bibinfo{pages}{043512} (\bibinfo{year}{1999}).

\bibitem[{\citenamefont{Yokoyama}(1999)}]{PTPS.136.338}
\bibinfo{author}{\bibfnamefont{J.}~\bibnamefont{Yokoyama}},
  \bibinfo{journal}{Prog. Theor. Phys. Suppl.} \textbf{\bibinfo{volume}{136}},
  \bibinfo{pages}{338} (\bibinfo{year}{1999}).

\bibitem[{\citenamefont{Saito et~al.}(2008)\citenamefont{Saito, Yokoyama, and
  Nagata}}]{1475-7516-2008-06-024}
\bibinfo{author}{\bibfnamefont{R.}~\bibnamefont{Saito}},
  \bibinfo{author}{\bibfnamefont{J.}~\bibnamefont{Yokoyama}}, \bibnamefont{and}
  \bibinfo{author}{\bibfnamefont{R.}~\bibnamefont{Nagata}},
  \bibinfo{journal}{J. Cosmol. Astropart. Phys.}
  \textbf{\bibinfo{volume}{2008}}, \bibinfo{pages}{024} (\bibinfo{year}{2008}).

\bibitem[{\citenamefont{Taruya}(1999)}]{PhysRevD.59.103505}
\bibinfo{author}{\bibfnamefont{A.}~\bibnamefont{Taruya}},
  \bibinfo{journal}{Phys. Rev. D} \textbf{\bibinfo{volume}{59}},
  \bibinfo{pages}{103505} (\bibinfo{year}{1999}).

\bibitem[{\citenamefont{Bassett and Tsujikawa}(2001)}]{PhysRevD.63.123503}
\bibinfo{author}{\bibfnamefont{B.~A.} \bibnamefont{Bassett}} \bibnamefont{and}
  \bibinfo{author}{\bibfnamefont{S.}~\bibnamefont{Tsujikawa}},
  \bibinfo{journal}{Phys. Rev. D} \textbf{\bibinfo{volume}{63}},
  \bibinfo{pages}{123503} (\bibinfo{year}{2001}).

\bibitem[{\citenamefont{Green and Malik}(2001)}]{PhysRevD.64.021301}
\bibinfo{author}{\bibfnamefont{A.~M.} \bibnamefont{Green}} \bibnamefont{and}
  \bibinfo{author}{\bibfnamefont{K.~A.} \bibnamefont{Malik}},
  \bibinfo{journal}{Phys. Rev. D} \textbf{\bibinfo{volume}{64}},
  \bibinfo{pages}{021301} (\bibinfo{year}{2001}).

\bibitem[{\citenamefont{Kawasaki et~al.}(2007)\citenamefont{Kawasaki, Takayama,
  Yamaguchi, and Yokoyama}}]{Kawasaki:2007zz}
\bibinfo{author}{\bibfnamefont{M.}~\bibnamefont{Kawasaki}},
  \bibinfo{author}{\bibfnamefont{T.}~\bibnamefont{Takayama}},
  \bibinfo{author}{\bibfnamefont{M.}~\bibnamefont{Yamaguchi}},
  \bibnamefont{and} \bibinfo{author}{\bibfnamefont{J.}~\bibnamefont{Yokoyama}},
  \bibinfo{journal}{Mod. Phys. Lett.} \textbf{\bibinfo{volume}{A22}},
  \bibinfo{pages}{1911} (\bibinfo{year}{2007}).

\bibitem[{\citenamefont{Kawaguchi et~al.}(2008)\citenamefont{Kawaguchi,
  Kawasaki, Takayama, Yamaguchi, and Yokoyama}}]{Kawaguchi:2007fz}
\bibinfo{author}{\bibfnamefont{T.}~\bibnamefont{Kawaguchi}},
  \bibinfo{author}{\bibfnamefont{M.}~\bibnamefont{Kawasaki}},
  \bibinfo{author}{\bibfnamefont{T.}~\bibnamefont{Takayama}},
  \bibinfo{author}{\bibfnamefont{M.}~\bibnamefont{Yamaguchi}},
  \bibnamefont{and} \bibinfo{author}{\bibfnamefont{J.}~\bibnamefont{Yokoyama}},
  \bibinfo{journal}{Mon. Not. Roy. Astron. Soc.}
  \textbf{\bibinfo{volume}{388}}, \bibinfo{pages}{1426} (\bibinfo{year}{2008}),
  \eprint{0711.3886}.

\bibitem[{\citenamefont{Bugaev and Klimai}(2009)}]{Bugaev:2008gw}
\bibinfo{author}{\bibfnamefont{E.}~\bibnamefont{Bugaev}} \bibnamefont{and}
  \bibinfo{author}{\bibfnamefont{P.}~\bibnamefont{Klimai}},
  \bibinfo{journal}{Phys.Rev.} \textbf{\bibinfo{volume}{D79}},
  \bibinfo{pages}{103511} (\bibinfo{year}{2009}), \eprint{0812.4247}.

\bibitem[{\citenamefont{Josan et~al.}(2009)\citenamefont{Josan, Green, and
  Malik}}]{Josan:2009qn}
\bibinfo{author}{\bibfnamefont{A.~S.} \bibnamefont{Josan}},
  \bibinfo{author}{\bibfnamefont{A.~M.} \bibnamefont{Green}}, \bibnamefont{and}
  \bibinfo{author}{\bibfnamefont{K.~A.} \bibnamefont{Malik}},
  \bibinfo{journal}{Phys.Rev.} \textbf{\bibinfo{volume}{D79}},
  \bibinfo{pages}{103520} (\bibinfo{year}{2009}), \eprint{0903.3184}.

\bibitem[{\citenamefont{Bringmann et~al.}(2012)\citenamefont{Bringmann, Scott,
  and Akrami}}]{Bringmann:2011ut}
\bibinfo{author}{\bibfnamefont{T.}~\bibnamefont{Bringmann}},
  \bibinfo{author}{\bibfnamefont{P.}~\bibnamefont{Scott}}, \bibnamefont{and}
  \bibinfo{author}{\bibfnamefont{Y.}~\bibnamefont{Akrami}},
  \bibinfo{journal}{Phys.Rev.} \textbf{\bibinfo{volume}{D85}},
  \bibinfo{pages}{125027} (\bibinfo{year}{2012}), \eprint{1110.2484}.

\bibitem[{\citenamefont{Josan and Green}(2010)}]{Josan:2010vn}
\bibinfo{author}{\bibfnamefont{A.~S.} \bibnamefont{Josan}} \bibnamefont{and}
  \bibinfo{author}{\bibfnamefont{A.~M.} \bibnamefont{Green}},
  \bibinfo{journal}{Phys.Rev.} \textbf{\bibinfo{volume}{D82}},
  \bibinfo{pages}{083527} (\bibinfo{year}{2010}), \eprint{1006.4970}.

\bibitem[{\citenamefont{Li et~al.}(2012)\citenamefont{Li, Erickcek, and
  Law}}]{Li:2012qha}
\bibinfo{author}{\bibfnamefont{F.}~\bibnamefont{Li}},
  \bibinfo{author}{\bibfnamefont{A.~L.} \bibnamefont{Erickcek}},
  \bibnamefont{and} \bibinfo{author}{\bibfnamefont{N.~M.} \bibnamefont{Law}},
  \bibinfo{journal}{Phys.Rev.} \textbf{\bibinfo{volume}{D86}},
  \bibinfo{pages}{043519} (\bibinfo{year}{2012}), \eprint{1202.1284}.

\bibitem[{\citenamefont{Scott et~al.}(2012)\citenamefont{Scott, Bringmann, and
  Akrami}}]{Scott:2012kx}
\bibinfo{author}{\bibfnamefont{P.}~\bibnamefont{Scott}},
  \bibinfo{author}{\bibfnamefont{T.}~\bibnamefont{Bringmann}},
  \bibnamefont{and} \bibinfo{author}{\bibfnamefont{Y.}~\bibnamefont{Akrami}},
  \bibinfo{journal}{J.Phys.Conf.Ser.} \textbf{\bibinfo{volume}{375}},
  \bibinfo{pages}{032012} (\bibinfo{year}{2012}), \eprint{1205.1432}.

\bibitem[{\citenamefont{Yang et~al.}(2013)\citenamefont{Yang, Yang, and
  Zong}}]{Yang:2013dsa}
\bibinfo{author}{\bibfnamefont{Y.}~\bibnamefont{Yang}},
  \bibinfo{author}{\bibfnamefont{G.}~\bibnamefont{Yang}}, \bibnamefont{and}
  \bibinfo{author}{\bibfnamefont{H.}~\bibnamefont{Zong}},
  \bibinfo{journal}{Phys.Rev.} \textbf{\bibinfo{volume}{D87}},
  \bibinfo{pages}{103525} (\bibinfo{year}{2013}), \eprint{1305.4213}.

\bibitem[{\citenamefont{Barrow and Coles}(1991)}]{Barrow2}
\bibinfo{author}{\bibfnamefont{J.}~\bibnamefont{Barrow}} \bibnamefont{and}
  \bibinfo{author}{\bibfnamefont{P.}~\bibnamefont{Coles}},
  \bibinfo{journal}{Mon. Not. Roy. Astron. Soc.}
  \textbf{\bibinfo{volume}{248}}, \bibinfo{pages}{52} (\bibinfo{year}{1991}).

\bibitem[{\citenamefont{Chluba and Sunyaev}(2011)}]{Chluba:2011hw}
\bibinfo{author}{\bibfnamefont{J.}~\bibnamefont{Chluba}} \bibnamefont{and}
  \bibinfo{author}{\bibfnamefont{R.}~\bibnamefont{Sunyaev}}
  (\bibinfo{year}{2011}), \eprint{1109.6552}.

\bibitem[{\citenamefont{Chluba et~al.}(2012{\natexlab{a}})\citenamefont{Chluba,
  Erickcek, and Ben-Dayan}}]{Chluba:2012we}
\bibinfo{author}{\bibfnamefont{J.}~\bibnamefont{Chluba}},
  \bibinfo{author}{\bibfnamefont{A.~L.} \bibnamefont{Erickcek}},
  \bibnamefont{and}
  \bibinfo{author}{\bibfnamefont{I.}~\bibnamefont{Ben-Dayan}},
  \bibinfo{journal}{Astrophys.J.} \textbf{\bibinfo{volume}{758}},
  \bibinfo{pages}{76} (\bibinfo{year}{2012}{\natexlab{a}}), \eprint{1203.2681}.

\bibitem[{\citenamefont{Chluba et~al.}(2012{\natexlab{b}})\citenamefont{Chluba,
  Khatri, and Sunyaev}}]{Chluba:2012gq}
\bibinfo{author}{\bibfnamefont{J.}~\bibnamefont{Chluba}},
  \bibinfo{author}{\bibfnamefont{R.}~\bibnamefont{Khatri}}, \bibnamefont{and}
  \bibinfo{author}{\bibfnamefont{R.~A.} \bibnamefont{Sunyaev}}
  (\bibinfo{year}{2012}{\natexlab{b}}), \eprint{1202.0057}.

\bibitem[{\citenamefont{Dent et~al.}(2012)\citenamefont{Dent, Easson, and
  Tashiro}}]{Dent:2012ne}
\bibinfo{author}{\bibfnamefont{J.~B.} \bibnamefont{Dent}},
  \bibinfo{author}{\bibfnamefont{D.~A.} \bibnamefont{Easson}},
  \bibnamefont{and} \bibinfo{author}{\bibfnamefont{H.}~\bibnamefont{Tashiro}},
  \bibinfo{journal}{Phys.Rev.} \textbf{\bibinfo{volume}{D86}},
  \bibinfo{pages}{023514} (\bibinfo{year}{2012}), \eprint{1202.6066}.

\bibitem[{\citenamefont{Khatri and Sunyaev}(2012)}]{Khatri:2012tw}
\bibinfo{author}{\bibfnamefont{R.}~\bibnamefont{Khatri}} \bibnamefont{and}
  \bibinfo{author}{\bibfnamefont{R.~A.} \bibnamefont{Sunyaev}},
  \bibinfo{journal}{JCAP} \textbf{\bibinfo{volume}{1209}}, \bibinfo{pages}{016}
  (\bibinfo{year}{2012}), \eprint{1207.6654}.

\bibitem[{\citenamefont{Sunyaev and Khatri}(2013)}]{Sunyaev:2013aoa}
\bibinfo{author}{\bibfnamefont{R.~A.} \bibnamefont{Sunyaev}} \bibnamefont{and}
  \bibinfo{author}{\bibfnamefont{R.}~\bibnamefont{Khatri}},
  \bibinfo{journal}{Int.J.Mod.Phys.} \textbf{\bibinfo{volume}{D22}},
  \bibinfo{pages}{1330014} (\bibinfo{year}{2013}), \eprint{1302.6553}.

\bibitem[{\citenamefont{Khatri and Sunyaev}(2013)}]{Khatri:2013dha}
\bibinfo{author}{\bibfnamefont{R.}~\bibnamefont{Khatri}} \bibnamefont{and}
  \bibinfo{author}{\bibfnamefont{R.~A.} \bibnamefont{Sunyaev}},
  \bibinfo{journal}{JCAP} \textbf{\bibinfo{volume}{1306}}, \bibinfo{pages}{026}
  (\bibinfo{year}{2013}), \eprint{1303.7212}.

\bibitem[{\citenamefont{Chluba and Grin}(2013)}]{Chluba:2013dna}
\bibinfo{author}{\bibfnamefont{J.}~\bibnamefont{Chluba}} \bibnamefont{and}
  \bibinfo{author}{\bibfnamefont{D.}~\bibnamefont{Grin}},
  \bibinfo{journal}{Mon.Not.Roy.Astron.Soc.} \textbf{\bibinfo{volume}{434}},
  \bibinfo{pages}{1619} (\bibinfo{year}{2013}), \eprint{1304.4596}.

\bibitem[{\citenamefont{Chluba and Jeong}(2013)}]{Chluba:2013pya}
\bibinfo{author}{\bibfnamefont{J.}~\bibnamefont{Chluba}} \bibnamefont{and}
  \bibinfo{author}{\bibfnamefont{D.}~\bibnamefont{Jeong}}
  (\bibinfo{year}{2013}), \eprint{1306.5751}.

\bibitem[{\citenamefont{Silk}(1968)}]{Silk:1967kq}
\bibinfo{author}{\bibfnamefont{J.}~\bibnamefont{Silk}},
  \bibinfo{journal}{Astrophys.J.} \textbf{\bibinfo{volume}{151}},
  \bibinfo{pages}{459} (\bibinfo{year}{1968}).

\bibitem[{\citenamefont{Sato}(1970)}]{HSato}
\bibinfo{author}{\bibfnamefont{H.}~\bibnamefont{Sato}}, \bibinfo{journal}{Prog.
  Theor. Phys.} \textbf{\bibinfo{volume}{45}}, \bibinfo{pages}{370}
  (\bibinfo{year}{1970}).

\bibitem[{\citenamefont{Sunyaev and Zeldovich}(1970)}]{Sunyaev:1970er}
\bibinfo{author}{\bibfnamefont{R.}~\bibnamefont{Sunyaev}} \bibnamefont{and}
  \bibinfo{author}{\bibfnamefont{Y.}~\bibnamefont{Zeldovich}},
  \bibinfo{journal}{Astrophys.Space Sci.} \textbf{\bibinfo{volume}{7}},
  \bibinfo{pages}{20} (\bibinfo{year}{1970}).

\bibitem[{\citenamefont{{Burigana} et~al.}(1991)\citenamefont{{Burigana},
  {Danese}, and {de Zotti}}}]{Burigana}
\bibinfo{author}{\bibfnamefont{C.}~\bibnamefont{{Burigana}}},
  \bibinfo{author}{\bibfnamefont{L.}~\bibnamefont{{Danese}}}, \bibnamefont{and}
  \bibinfo{author}{\bibfnamefont{G.}~\bibnamefont{{de Zotti}}},
  \bibinfo{journal}{Astron. Astrophys.} \textbf{\bibinfo{volume}{246}},
  \bibinfo{pages}{49} (\bibinfo{year}{1991}).

\bibitem[{\citenamefont{Hu and Silk}(1993)}]{Hu:1992dc}
\bibinfo{author}{\bibfnamefont{W.}~\bibnamefont{Hu}} \bibnamefont{and}
  \bibinfo{author}{\bibfnamefont{J.}~\bibnamefont{Silk}},
  \bibinfo{journal}{Phys.Rev.} \textbf{\bibinfo{volume}{D48}},
  \bibinfo{pages}{485} (\bibinfo{year}{1993}).

\bibitem[{\citenamefont{Steigman and Nollett}(2014)}]{Steigman:2014pfa}
\bibinfo{author}{\bibfnamefont{G.}~\bibnamefont{Steigman}} \bibnamefont{and}
  \bibinfo{author}{\bibfnamefont{K.~M.} \bibnamefont{Nollett}}
  (\bibinfo{year}{2014}), \eprint{1401.5488}.

\bibitem[{\citenamefont{Ade et~al.}(2013{\natexlab{b}})}]{Ade:2013zuv}
\bibinfo{author}{\bibfnamefont{P.}~\bibnamefont{Ade}} \bibnamefont{et~al.}
  (\bibinfo{collaboration}{Planck Collaboration})
  (\bibinfo{year}{2013}{\natexlab{b}}), \eprint{1303.5076}.

\bibitem[{\citenamefont{Simha and Steigman}(2008)}]{Simha:2008zj}
\bibinfo{author}{\bibfnamefont{V.}~\bibnamefont{Simha}} \bibnamefont{and}
  \bibinfo{author}{\bibfnamefont{G.}~\bibnamefont{Steigman}},
  \bibinfo{journal}{JCAP} \textbf{\bibinfo{volume}{0806}}, \bibinfo{pages}{016}
  (\bibinfo{year}{2008}), \eprint{0803.3465}.

\bibitem[{\citenamefont{Barrow}(1977)}]{Barrow1}
\bibinfo{author}{\bibfnamefont{J.}~\bibnamefont{Barrow}},
  \bibinfo{journal}{Nature} \textbf{\bibinfo{volume}{267}},
  \bibinfo{pages}{117} (\bibinfo{year}{1977}).

\bibitem[{\citenamefont{Jedamzik et~al.}(1998)\citenamefont{Jedamzik,
  Katalinic, and Olinto}}]{Jedamzik:1996wp}
\bibinfo{author}{\bibfnamefont{K.}~\bibnamefont{Jedamzik}},
  \bibinfo{author}{\bibfnamefont{V.}~\bibnamefont{Katalinic}},
  \bibnamefont{and} \bibinfo{author}{\bibfnamefont{A.~V.}
  \bibnamefont{Olinto}}, \bibinfo{journal}{Phys.Rev.}
  \textbf{\bibinfo{volume}{D57}}, \bibinfo{pages}{3264} (\bibinfo{year}{1998}),
  \eprint{astro-ph/9606080}.

\bibitem[{\citenamefont{Saito and Yokoyama}(2009)}]{Saito:2008jc}
\bibinfo{author}{\bibfnamefont{R.}~\bibnamefont{Saito}} \bibnamefont{and}
  \bibinfo{author}{\bibfnamefont{J.}~\bibnamefont{Yokoyama}},
  \bibinfo{journal}{Phys. Rev. Lett.} \textbf{\bibinfo{volume}{102}},
  \bibinfo{pages}{161101} (\bibinfo{year}{2009}), \bibinfo{note}{[{\bf 107},
  069901(E) (2011).]}, \eprint{0812.4339}.

\bibitem[{\citenamefont{Saito and Yokoyama}(2010)}]{Saito:2009jt}
\bibinfo{author}{\bibfnamefont{R.}~\bibnamefont{Saito}} \bibnamefont{and}
  \bibinfo{author}{\bibfnamefont{J.}~\bibnamefont{Yokoyama}},
  \bibinfo{journal}{Prog. Theor. Phys.} \textbf{\bibinfo{volume}{123}},
  \bibinfo{pages}{867} (\bibinfo{year}{2010}), \bibinfo{note}{[{\bf 126},
  351(E) (2011).]}, \eprint{0912.5317}.

\bibitem[{\citenamefont{Carr et~al.}(2010)\citenamefont{Carr, Kohri, Sendouda,
  and Yokoyama}}]{Carr:2009jm}
\bibinfo{author}{\bibfnamefont{B.}~\bibnamefont{Carr}},
  \bibinfo{author}{\bibfnamefont{K.}~\bibnamefont{Kohri}},
  \bibinfo{author}{\bibfnamefont{Y.}~\bibnamefont{Sendouda}}, \bibnamefont{and}
  \bibinfo{author}{\bibfnamefont{J.}~\bibnamefont{Yokoyama}},
  \bibinfo{journal}{Phys. Rev. D} \textbf{\bibinfo{volume}{81}},
  \bibinfo{pages}{104019} (\bibinfo{year}{2010}), \eprint{0912.5297}.

\bibitem[{\citenamefont{Niemeyer and Jedamzik}(1999)}]{PhysRevD.59.124013}
\bibinfo{author}{\bibfnamefont{J.~C.} \bibnamefont{Niemeyer}} \bibnamefont{and}
  \bibinfo{author}{\bibfnamefont{K.}~\bibnamefont{Jedamzik}},
  \bibinfo{journal}{Phys. Rev. D} \textbf{\bibinfo{volume}{59}},
  \bibinfo{pages}{124013} (\bibinfo{year}{1999}).

\bibitem[{\citenamefont{Musco et~al.}(2009)\citenamefont{Musco, Miller, and
  Polnarev}}]{Musco:2008hv}
\bibinfo{author}{\bibfnamefont{I.}~\bibnamefont{Musco}},
  \bibinfo{author}{\bibfnamefont{J.~C.} \bibnamefont{Miller}},
  \bibnamefont{and} \bibinfo{author}{\bibfnamefont{A.~G.}
  \bibnamefont{Polnarev}}, \bibinfo{journal}{Class. Quant. Grav.}
  \textbf{\bibinfo{volume}{26}}, \bibinfo{pages}{235001}
  (\bibinfo{year}{2009}), \eprint{0811.1452}.

\bibitem[{\citenamefont{Byrnes et~al.}(2012)\citenamefont{Byrnes, Copeland, and
  Green}}]{Byrnes:2012yx}
\bibinfo{author}{\bibfnamefont{C.~T.} \bibnamefont{Byrnes}},
  \bibinfo{author}{\bibfnamefont{E.~J.} \bibnamefont{Copeland}},
  \bibnamefont{and} \bibinfo{author}{\bibfnamefont{A.~M.} \bibnamefont{Green}},
  \bibinfo{journal}{Phys.Rev.} \textbf{\bibinfo{volume}{D86}},
  \bibinfo{pages}{043512} (\bibinfo{year}{2012}), \eprint{1206.4188}.

\bibitem[{\citenamefont{Young and Byrnes}(2013)}]{Young:2013oia}
\bibinfo{author}{\bibfnamefont{S.}~\bibnamefont{Young}} \bibnamefont{and}
  \bibinfo{author}{\bibfnamefont{C.~T.} \bibnamefont{Byrnes}},
  \bibinfo{journal}{JCAP} \textbf{\bibinfo{volume}{1308}}, \bibinfo{pages}{052}
  (\bibinfo{year}{2013}), \eprint{1307.4995}.

\bibitem[{\citenamefont{Kohri et~al.}(2014)\citenamefont{Kohri, Suyama, and
  Nakama}}]{KNS}
\bibinfo{author}{\bibfnamefont{K.}~\bibnamefont{Kohri}},
  \bibinfo{author}{\bibfnamefont{T.}~\bibnamefont{Suyama}}, \bibnamefont{and}
  \bibinfo{author}{\bibfnamefont{T.}~\bibnamefont{Nakama}},
  \bibinfo{journal}{in prep.}  (\bibinfo{year}{2014}).

\bibitem[{\citenamefont{Jeong et~al.}(2014)\citenamefont{Jeong, Pradler,
  Chluba, and Kamionkowski}}]{Jeong:2014gna}
\bibinfo{author}{\bibfnamefont{D.}~\bibnamefont{Jeong}},
  \bibinfo{author}{\bibfnamefont{J.}~\bibnamefont{Pradler}},
  \bibinfo{author}{\bibfnamefont{J.}~\bibnamefont{Chluba}}, \bibnamefont{and}
  \bibinfo{author}{\bibfnamefont{M.}~\bibnamefont{Kamionkowski}}
  (\bibinfo{year}{2014}), \eprint{1403.3697}.

\end{thebibliography}

\end{document}